\documentclass[12pt]{article}

\usepackage{amssymb, amsthm, amsmath, amsfonts, eurosym, newtxmath, nccmath, bm, esvect, mdframed}

\usepackage{graphicx, subfig}

\usepackage{booktabs, dcolumn, float, adjustbox, multirow, threeparttable, siunitx}

\usepackage{hyperref}
\usepackage[font = small]{caption}

\usepackage[textsize = small]{todonotes}

\usepackage[backend = biber, style = apa, uniquename = false, sorting = ynt]{biblatex}

\usepackage{geometry, color, comment, setspace, pdflscape, indentfirst}
\usepackage[bottom]{footmisc}

\usepackage{hyperref}
 \hypersetup{
     colorlinks=true,    
     citecolor=blue,
     filecolor=magenta,  
     linkcolor=blue,
     urlcolor=blue,
}
\newcolumntype{L}[1]{>{\raggedright\let\newline\\arraybackslash\hspace{0pt}}m{#1}}
\newcolumntype{C}[1]{>{\centering\let\newline\\arraybackslash\hspace{0pt}}m{#1}}
\newcolumntype{R}[1]{>{\raggedleft\let\newline\\arraybackslash\hspace{0pt}}m{#1}}

\definecolor{linkcolour}{rgb}{0, 0.2, 0.6}
\hypersetup{colorlinks, breaklinks, urlcolor = linkcolour, linkcolor = linkcolour, citecolor = linkcolour}

\theoremstyle{definition}
\newtheorem{assumption}{Assumption}

\newtheoremstyle{remark}
  {3pt}                     % Space above
  {3pt}                     % Space below
  {}                        % Body font (empty means default)
  {}                        % Indent amount
  {\bfseries}               % Theorem head font
  {.}                       % Punctuation after theorem head
  { }                       % Space after theorem head
  {}                        % Theorem head spec (empty means normal)

\theoremstyle{remark}

\newmdtheoremenv{remarkk}{Remark}

\newcounter{assumptionAlternative}

\newtheorem{altAssumptionEnv}[assumptionAlternative]{Assumption}

\usepackage[title]{appendix}

\usepackage[ruled]{algorithm2e}

\DeclareMathOperator{\EX}{\mathbb{E}} % Expected value.
\DeclareMathOperator{\PX}{\mathbb{P}} % Probability.
\newcommand{\indep}{\perp \!\!\! \perp} % Independence.
\newcommand{\indicator}[1]{\vmathbb{1} ( #1 )} % Indicator function.

\def\vector#1{\mbox{\boldmath{$#1$}}} % Vectors.
\newcommand{\open}{“} 

\DeclareFontFamily{U}{mathx}{}
\DeclareFontShape{U}{mathx}{m}{n}{<-> mathx10}{}
\DeclareSymbolFont{mathx}{U}{mathx}{m}{n}
\DeclareMathAccent{\widehat}{0}{mathx}{"70}
\DeclareMathAccent{\widecheck}{0}{mathx}{"71}

\bibliography{references.bib}

\newcommand{\outcome}[0]{Y_{i}} 
\newcommand{\outcomeTime}[1]{Y_{i#1}} 
\newcommand{\potentialOutcome}[1]{Y_{i} ( #1 )} 
\newcommand{\potentialOutcomeTime}[2]{Y_{i#2} ( #1 )} 
\newcommand{\potentialOutcomeIV}[2]{Y_{i} ( #1, #2 )}

\newcommand{\propensityScore}[1]{\pi ( #1 )} 
\newcommand{\estimatedPropensityScore}[1]{\hat{\pi} ( #1 )} 
\newcommand{\drScores}[1]{\Gamma_{#1i}}
\newcommand{\estimatedDrScores}[1]{\widehat{\Gamma}_{#1i}}

\newcommand{\conditionalClassProbability}[3]{p_{#1} ( #2, #3 )} 
 
\newcommand{\probabilityShift}[1]{\delta_{#1}} 
\newcommand{\probabilityShiftTreated}[1]{\delta_{#1, T}} 
\newcommand{\LocalProbabilityShift}[1]{\delta_{#1, L}} 
\newcommand{\conditionalProbabilityShift}[2]{\delta_{#1} ( #2 )} 
\newcommand{\probabilityShiftCutoff}[1]{\delta_{#1, C}}

\newcommand{\estimatedConditionalClassProbability}[3]{\hat{p}_{#1} ( #2, #3 )} 
\newcommand{\estimatedProbabilityShift}[1]{\hat{\delta}_{#1}} 
\newcommand{\estimatedProbabilityShiftTreated}[1]{\hat{\delta}_{#1, T}}

\begin{document}

%%%%%%%%%%%%%%%%%%%%%%%%%%%%%%%%%%%%%%%%
%            TITLE PAGE.               %
%%%%%%%%%%%%%%%%%%%%%%%%%%%%%%%%%%%%%%%%

\begin{titlepage}
    \title{Causal Inference for Qualitative Outcomes}
    
    \author{Riccardo Di Francesco\thanks{\ Department of Economics, University of Southern Denmark, Denmark, email: \href{mailto:rdif@sam.sdu.dk}{rdif@sam.sdu.dk}.}
    \and 
    Giovanni Mellace\thanks{\ Department of Economics, University of Southern Denmark, Denmark, email: \href{mailto:giome@sam.sdu.dk}{giome@sam.sdu.dk}.}}
    
    \date{\today}
    
    \maketitle

\begin{abstract}
    \noindent Causal inference methods such as instrumental variables, regression discontinuity, and difference-in-differences are widely used to identify and estimate treatment effects. However, when outcomes are qualitative, their application poses fundamental challenges. This paper highlights these challenges and proposes an alternative framework that focuses on well-defined and interpretable estimands. We show that conventional identification assumptions suffice for identifying the new estimands and outline simple, intuitive estimation strategies that remain fully compatible with conventional econometric methods. We provide an accompanying open-source R package, \texttt{causalQual}, which is publicly available on CRAN.  
    \end{abstract}
    
    \setcounter{page}{0}
    
    \thispagestyle{empty}
\end{titlepage}

\pagebreak \newpage

\doublespacing

%%%%%%%%%%%%%%%%%%%%%%%%%%%%%%%%%%%%%%%%
%            INTRODUCTION.             %
%%%%%%%%%%%%%%%%%%%%%%%%%%%%%%%%%%%%%%%%

\section{Introduction}
\label{sec_introduction}
\noindent Causal inference has become a cornerstone of applied econometrics, with methods such as \textit{instrumental variables}, \textit{difference-in-differences}, and \textit{regression discontinuity} widely employed to identify and estimate the effects of policies or interventions \parencite{imbens2022causality, angrist2022empirical}.

These methodologies are primarily developed for numerical outcomes, whether continuous or discrete. Yet, many empirical applications across disciplines involve qualitative outcomes, which take the form of ordered or unordered categories.\footnote{\ In political science, voter ideology is often recorded on an ordinal scale ranging from \open very liberal" to \open very conservative" \parencite{mason2015disrespectfully}, and attitudes toward policy issues are commonly measured using Likert scales \parencite{kuziemko2015elastic}. In health, researchers analyze life satisfaction \parencite{frey2002can}, subjective well-being \parencite{reisinger2022subjective}, and self-reported health status \parencite{peracchi2012heterogeneity, peracchi2013heterogeneous}---all of which lack a natural cardinal interpretation. In consumer research, perceived product quality and satisfaction are often captured through consumer ratings and review scores \parencite{mayzlin2014promotional}, which are inherently ordinal.} In such cases, widely used causal estimands, such as the \textit{average treatment effect} and its conditional versions, are ill-defined, as arithmetic operations on outcome categories lack meaningful interpretation.

% However, these methods are primarily designed for numerical outcomes. Many empirical applications, from political science to health and consumer research, involve qualitative outcomes such as voter ideology \parencite{mason2015disrespectfully}, life satisfaction \parencite{frey2002can}, and product ratings \parencite{mayzlin2014promotional}. In such contexts, standard causal estimands are ill-defined as arithmetic operations on categories lack meaningful interpretation.

This paper highlights the fundamental challenges that arise when conventional causal inference methods are applied to qualitative outcomes and introduces a principled approach to addressing these issues. A common approach among applied researchers is to dichotomize the outcome based on an arbitrary threshold. However, results may be highly sensitive to the choice of threshold, and the validity of identifying assumptions for one dichotomization does not necessarily imply identification under an alternative thresholding scheme \parencite{yamauchi2020difference}. We advocate for a fundamental shift toward alternative estimands that are meaningful, interpretable, and identifiable within standard research designs.

% A common, yet problematic, approach is to dichotomize outcomes, which can lead to results highly sensitive to arbitrary threshold choices, and the validity of identifying assumptions for one dichotomization does not necessarily imply identification under an different thresholding scheme \parencite{yamauchi2020difference}. We advocate for a fundamental shift towards alternative estimands that are meaningful, interpretable, and identifiable within standard research designs.

Specifically, we demonstrate that a naive application of widely used research designs to qualitative outcomes leads to the identification of quantities whose magnitude and sign depend on arbitrary coding, rendering them economically meaningless. We thus shift our focus to well-defined and interpratable estimands that characterize how treatment affects the probability distribution over outcome categories. We establish that conventional identification assumptions---or slight modifications thereof---suffice for identifying these new estimands, ensuring compatibility with standard research designs. For estimation, we rely on conventional econometric methods, but explicitly model category probabilities. We further provide an accompanying open-source R package to facilitate implementation---named \texttt{causalQual}.\footnote{\  Our framework, though introduced with basic examples, naturally extends to more complex settings,  including instrumental variables with multiple instruments \parencite[see, e.g.,][]{HLM2025,mogstad2018using,mogstad2018identification,goff2024vector}, difference-in-differences with heterogeneous effects and staggered treatment adoption \parencite[see, e.g.,][]{callaway2021difference,SunAbraham2021,roth2023s, deChaisemartin2023,Borusyak2024, goodman2021difference}, and fuzzy regression discontinuity designs  \parencite[see, e.g.,][]{cattaneo2022regression}. We plan to continuously update our R package to incorporate these extensions as they develop. Users interested in specific extensions are encouraged to contribute by submitting feature requests or discussions at \href{https://github.com/riccardo-df/causalQual/issues}{https://github.com/riccardo-df/causalQual/issues}.}

% Our framework, though shown with basic examples, supports advanced cases like instrumental variables with multiple instruments, heterogeneous difference-in-differences, and fuzzy regression discontinuity designs. We'll add these extensions to our R package over time. To request features or contribute, visit our GitHub: 

While the issues we highlight may seem straightforward, we note that causal inference with qualitative outcomes remains underdeveloped. Existing research proposes estimands---such
as the probabilities that treatment is (strictly) beneficial---that depend on the joint distribution of potential outcomes, whose identification is pursed under restrictive conditions \parencite{volfovsky2015causal, agresti2017ordinal, chiba2018bayesian}.\footnote{\ \textcite{lu2018treatment, lu2020sharp} adopt a partial identification approach, deriving sharp bounds for these estimands using only the marginal distributions of potential outcomes.} In contrast, we introduce a class of estimands that apply to both multinomial
and ordered outcomes and achieve identification without requiring additional assumptions
beyond those conventionally made in empirical research.\footnote{\  \textcite{yamauchi2020difference} studies a related estimand within a difference-in-differences framework, imposing strong parametric and distributional assumptions on the outcome as well as a distributional parallel trends assumption \parencite{athey2006identification}.}

%The rest of the paper is organized as follows. Section \ref{sec_causal_framework} introduces the causal framework. Section \ref{sec_selection_on_obseravbles}, \ref{sec_instrumental_variables}, \ref{sec_regression_discontinuity}, and \ref{sec_difference_in_differences} discuss identification and estimation under selection-on-obseravbles, instrumental variables, difference-in-differences, and regression discontinuity designs, respectively. Section \ref{sec_simulation} presents simulation results. Section \ref{sec_conclusion} concludes.

%%%%%%%%%%%%%%%%%%%%%%%%%%%%%%%%%%%%%%%%
%          CAUSAL FRAMEWORK.           %
%%%%%%%%%%%%%%%%%%%%%%%%%%%%%%%%%%%%%%%%

\section{Causal framework}\label{sec_causal_framework}
\noindent We study the causal effect of a binary treatment $D_i \in \{0, 1 \}$ on a qualitative---either multinomial or ordered---outcome $\outcome \in \{ 1, \dots, M \}$, with $\vector X_i$ as pre-treatment covariates. We adopt the potential outcomes framework \parencite{neyman1923, rubin1974estimating} and use $\potentialOutcome{1}$ and $\potentialOutcome{0}$ to denote the outcome that unit $i$ would experience under each treatment level, with $\outcome = D_i \potentialOutcome{1} + ( 1 - D_i ) \potentialOutcome{0}$. We further define $\conditionalClassProbability{m}{d}{\vector x} := \PX ( \outcome = m | D_i = d, \vector X_i = \vector x )$ as the conditional class probability and $\propensityScore{\vector x} := \PX ( D_i = 1 | \vector X_i = \vector x )$ as the propensity score. 

%Since $\outcome$ is qualitative, standard arithmetic operations (e.g., averages) lack meaning.
% Adopting the potential outcomes framework \parencite{neyman1923, rubin1974estimating}, we denote as $\potentialOutcome{1}$ and $\potentialOutcome{0}$ as outcomes under each treatment level, with $\outcome = D_i \potentialOutcome{1} + ( 1 - D_i ) \potentialOutcome{0}$. While a cardinal outcome would allow for an individual treatment effect $\potentialOutcome{1} - \potentialOutcome{0}$ and well-defined ATE/ATT, these are meaningless for qualitative variables.

If $\outcome$ were a cardinal numeric variable, the individual treatment effect could be naturally defined as $\potentialOutcome{1} - \potentialOutcome{0}$. In that case, standard causal estimands such as the Average Treatment Effect (ATE) $\tau := \EX [ \potentialOutcome{1} - \potentialOutcome{0}]$ and the Average Treatment Effect on the Treated (ATT)  $\tau_T := \EX [ \potentialOutcome{1} - \potentialOutcome{0} | D_i = 1]$ would be well-defined and directly identifiable under additional assumptions. 

However, because $\outcome$ is a qualitative variable, such differences are not meaningful. Instead, we define
\begin{equation}
    \begin{split}
        \probabilityShift{m} := \PX ( \potentialOutcome{1} = m ) - \PX ( \potentialOutcome{0} = m )
    \end{split}
    \label{equation_probability_shift}
\end{equation}
as the Probability of Shift (PS)---the change in the probability of belonging to category $m$ due to treatment.\footnote{\ For instance, $\probabilityShift{m} > 0$ indicates that the treatment increases the likelihood of observing outcome category $m$. By construction, $\sum_{m = 1}^M \probabilityShift{m} = 0$, reflecting the intutitive trade-off that an increase in the probability of some outcome categories must be offset by a decrease in others.} We further define two conditional versions of PS: the Probability of Shift on the Treated (PST)
\begin{equation}
    \probabilityShiftTreated{m} := \PX ( \potentialOutcome{1} = m | D_i = 1 ) - \PX ( \potentialOutcome{0} = m | D_i = 1 ),
\end{equation}
and the Conditional Probability of Shift (CPS)
\begin{equation}
    \conditionalProbabilityShift{m}{\vector x} := \PX ( \potentialOutcome{1} = m | \vector X_i = \vector x ) - \PX ( \potentialOutcome{0} = m | \vector X_i = \vector x ).
\end{equation}

\begin{remarkk}
    \label{remark_ate}
    A naive approach is to treat $\outcome$ as numeric, in which case the ATE would correspond to a weighted sum of $\probabilityShift{m}$ with weights given by category labels $m$:
    \begin{equation}
        \tau = \sum_{m = 1}^M m \probabilityShift{m}.
    \end{equation}
    Any attempt at interpreting $\tau$ is fundamentally flawed for at least two reasons. First, the weights $m$ are an artifact of arbitrary labeling and they can be reindexed without altering the underlying outcome---yet doing so would completely change the magnitude of the ATE. This lack of invariance makes the ATE highly sensitive to arbitrary coding choices.
    Second, since $\sum_{m = 1}^M \probabilityShift{m} = 0$, any nonzero treatment effect necessarily implies $\probabilityShift{m} > 0$ for some categories and $\probabilityShift{m} < 0$ for others. Consequently, small shifts in higher-indexed categories may dominate larger shifts in lower-indexed ones, leading to an unreliable and misleading sign.    
\end{remarkk}

%%%%%%%%%%%%%%%%%%%%%%%%%%%%%%%%%%%%%%%%
%           IDENTIFICATION.            %
%%%%%%%%%%%%%%%%%%%%%%%%%%%%%%%%%%%%%%%%

\section{Identification under different designs}
\noindent In this section, we discuss the identification of PS and conditional versions thereof across several commonly used research designs. We conclude with a brief discussion on estimation strategies. Identification proofs are provided in Appendix \ref{app_proofs}, and simulation results are provided in Appendix \ref{app_simulation}.

%%% SELECTION-ON-OBSERVABLES.
\subsection{Selection-on-observables} 
\label{sec_selection_on_obseravbles}
\noindent Selection-on-observables research designs are based on the following standard assumptions \parencite[see, e.g.,][]{imbens2015causal}:
\begin{assumption}
    (Unconfoundedness): $\{ \potentialOutcome{1}, \potentialOutcome{0} \} \indep D_i | \vector X_i$.
    \label{assumption_unconfoundedness}
\end{assumption}
\begin{assumption}
    (Common support): $0 < \propensityScore{\vector X_i} < 1$.
    \label{assumption_common_support}
\end{assumption}
%
%\noindent Assumption \ref{assumption_unconfoundedness} states that, after conditioning on $\vector X_i$, treatment assignment is as good as random, meaning that $\vector X_i$ fully accounts for selection into treatment. Assumption \ref{assumption_common_support} ensures that for every value of $\vector X_i$, there are both treated and untreated units, preventing situations where no valid counterfactual exists. 

\noindent Together, these assumptions allow for the identification of $\tau$ and conditional versions thereof. However, the qualitative nature of $\outcome$ renders $\tau$ ill-defined. Therefore, we shift our focus to the identification of PS and its conditional counterparts.

Specifically, under Assumptions \ref{assumption_unconfoundedness}--\ref{assumption_common_support},  
\begin{equation*}
    \begin{split}
        \conditionalProbabilityShift{m}{\vector X_i} = \conditionalClassProbability{m}{1}{\vector X_i} & - \conditionalClassProbability{m}{0}{\vector X_i}.
    \end{split}
    \label{equation_identification}
\end{equation*}
This establishes that the standard selection-on-observables assumptions are sufficient to identify the CPS. Moreover, since $\probabilityShift{m} = \EX [ \conditionalProbabilityShift{m}{\vector X_i} ]$ and $\probabilityShiftTreated{m} = \EX [ \conditionalProbabilityShift{m}{\vector X_i} | D_i = 1 ]$, the PS and PST are also identified.% Thus, even though standard causal parameters are ill-defined for qualitative outcomes, selection-on-observables remains a valid identification strategy. The key insight is to shift the focus toward meaningful and interpretable estimands, namely the PS and its conditional versions.

% For estimation, note that
% %
% \begin{equation}
%     \begin{split}
%         \probabilityShift{m} = \EX[ \EX [ \indicator{\outcome = m} | D_i = 1, \vector X_i = \vector x ] - \EX [ \indicator{\outcome = m} | D_i = 0, \vector X_i = \vector x ]].
%     \end{split}
% \end{equation}
% %
% This suggests that 

% We can construct an estimate $\estimatedProbabilityShift{m}$ ($\estimatedProbabilityShiftTreated{m}$) by applying standard estimation strategies to the binary variable $\indicator{\outcome = m}$ and estimating the ATE (ATT) on that variable.%\footnote{\ Applying such estimation strategies to the subsample of treated units would yield an estimate $\estimatedProbabilityShiftTreated{m}$ of $\probabilityShiftTreated{m}$.}

%%% INSTRUMENTAL VARIABLES.
\subsection{Instrumental variables} 
\label{sec_instrumental_variables}
% \noindent %In some applications, the observed covariates $\vector X_i$ may not fully account for selection into treatment, making a selection-on-observables design unsuitable. Instrumental variables (IV) designs provide a popular identification strategy to address this issue. The key idea behind IV is to exploit a variable -- an instrument -- that is as good as randomly assigned, influences treatment assignment, but has no direct effect on the outcome except through treatment. This exogenous source of variation enables the identification of causal effects by isolating changes in treatment status that are not driven by unobserved factors.
% Although our results can be readily extended to more complex settings, we focus on the simple case of a single binary instrument $Z_i \in \{ 0, 1 \}$ for ease of illustration. We define the potential treatment statuses $D_i(1)$ and $D_i(0)$, which represent the treatment that unit $i$ would receive under each level of the instrument. Potential outcomes are now indexed by both treatment status $d$ and instrument assignment $z$, denoted by $\potentialOutcomeIV{d}{z}$.%\footnote{\ A typical example arises in experimental settings where $Z_i$ represents randomized treatment assignment, but not all units comply with their assigned treatment—meaning that for some units, $D_i \neq Z_i$.}
\noindent We focus on the simple case of a single binary instrument $Z_i \in \{ 0, 1 \}$ for ease of illustration. We use $D_i(1)$ and $D_i(0)$ to define the potential treatment statuses and $\potentialOutcomeIV{d}{z}$ to index potential outcomes by both treatment status $d$ and instrument assignment $z$. We classify units into four distinct compliance types \parencite{angrist1996identification}: \textit{always-takers} ($at$) ($D_i(1) = D_i(0) = 1$), \textit{never-takers} ($nt$)  ($D_i(1) = D_i(0) = 0$), \textit{compliers} ($co$) ($D_i(1) = 1, D_i(0) = 0$), and \textit{defiers} ($de$) ($D_i(1) = 0, D_i(0) = 1$). To shorten notation, we denote an unit's compliance type by $T_i \in \{at, nt, co, de\}$.

% Under this framework, units can be classified into four distinct compliance types based on their potential treatment status as a function of the instrument \parencite{angrist1996identification}: always takers (AT), who receive treatment regardless of assignment ($D_i(1) = D_i(0) = 1$); never takers (NT), who never receive treatment ($D_i(1) = D_i(0) = 0$); compliers (CO), who take treatment only if assigned ($D_i(1) = 1, D_i(0) = 0$); and defiers (DE), who take treatment only when not assigned ($D_i(1) = 0, D_i(0) = 1$). We denote an unit's compliance type by $T_i \in \{\text{AT}, \text{NT}, \text{CO}, \text{DE}\}$.

A standard approach in Instrumental Variables (IV) settings is to target identification and estimation of the Local Average Treatment Effect (LATE) $\tau_L := \EX [ \potentialOutcome{1} - \potentialOutcome{0} | T_i=co]$. Identification relies on the following standard assumptions \parencite{imbens1994identification, angrist1996identification}:
\begin{assumption} 
    (Exogeneity): $\{ \potentialOutcomeIV{D_i}{1}, \potentialOutcomeIV{D_i}{0}, D_i(1), D_i(0) \} \indep Z_i$. 
    \label{assumption_exogeneity_iv} 
\end{assumption} 
\begin{assumption} 
    (Exclusion restriction): $\potentialOutcomeIV{d}{1} = \potentialOutcomeIV{d}{0} = \potentialOutcome{d}, \, d = 0, 1$. 
    \label{assumption_exclusion} 
\end{assumption} 
\begin{assumption} 
    (Monotonicity): $\PX (D_i(1) \geq D_i(0))=1$. 
    \label{assumption_monotonicity} 
\end{assumption}
\begin{assumption} 
    (Relevance): $\PX(T_i=co) > 0$. 
    \label{assumption_relevance} 
\end{assumption} 
% 

%\noindent Assumption \ref{assumption_exogeneity_iv} mandates that the instrument is as good as randomly assigned. Assumption \ref{assumption_exclusion} requires that the instrument affects the outcome only through its influence on treatment, ruling out any direct effect. Assumption \ref{assumption_monotonicity} imposes that the instrument can only increase the likelihood of treatment, ruling out defiers.\footnote{\ This assumption is made without loss of generality; one could alternatively assume that the instrument can only decrease the probability of treatment.} Finally, Assumption \ref{assumption_relevance} states that the instrument has a nonzero effect on the treatment, thereby generating exogenous variation in the latter.

However, the qualitative nature of $\outcome$ renders $\tau_L$ ill-defined. We thus shift our focus toward the identification of the Local Probability Shift (LPS):
\begin{equation}
    \LocalProbabilityShift{m} := \PX ( \potentialOutcome{1} = m |T_i=co ) - \PX ( \potentialOutcome{0} = m | T_i=co).
\end{equation}
Specifically, Appendix \ref{app_proofs} shows that, under Assumptions \ref{assumption_exogeneity_iv}--\ref{assumption_relevance},
\begin{equation}
    \LocalProbabilityShift{m} = \frac{\PX ( \outcome = m |Z_i=1 )-\PX ( \outcome = m |Z_i=0 )}{\PX(D_i=1|Z_i=1 )-\PX(D_i=1|Z_i=0)}.
\end{equation}
This establishes that the standard IV assumptions are sufficient to identify the LPS. 

%\footnote{\ Unlike in the selection-on-observables setting, estimation remains the same regardless of whether $\outcome$ is multinomial or ordered.} 

%Specifically, consider the following model: 
% % 
% \begin{align} 
%     D_i & = \gamma_0 + \gamma_1 Z_i + \nu_i, 
%     \label{equation_first_stage_iv} \\ 
%     \indicator{\outcome = m} & = \alpha_{m0} + \alpha_{m1} \widehat{D}_i + \epsilon_{mi},
%     \label{equation_second_stage_iv} 
% \end{align} 
% 
%with $\widehat{D}_i$ the predicted values constructed by fitting the first-stage regression model \eqref{equation_first_stage_iv} via OLS. A well-established result in the IV literature is that, under Assumptions \ref{assumption_exogeneity_iv}--\ref{assumption_monotonicity}, $\alpha_{m1} = \LocalProbabilityShift{m}$. Therefore, one can estimate the second-stage regression model \eqref{equation_second_stage_iv} via OLS and use the resulting estimate $\hat{\alpha}_{m1}$ as an estimate of $\LocalProbabilityShift{m}$ -- a procedure known as two-stage least squares.

%%% REGRESSION DISCONTINUITY.
\subsection{Regression discontinuity} 
\label{sec_regression_discontinuity}
%Regression Discontinuity (RD) designs are employed to identify causal effects when treatment assignment is determined by whether a continuous variable crosses a known threshold or cutoff. The key assumption is that units just above and just below the cutoff are comparable, meaning that any discontinuity in outcomes at the threshold can be attributed to the treatment rather than to pre-existing differences \parencite[see, e.g.,][]{lee2010regression}.
\noindent  We focus on the standard \textit{sharp} Regression Discontinuity (RD) design for ease of illustration. In this setup, we consider a single observed covariate $X_i$---the \open running variable." Units receive treatment if their running variable exceeds some threshold $c$ ($D_i = \indicator{X_i \geq c}$). %Thus, $D_i = \indicator{X_i \geq c}$. 

%This setup leads to two key observations that draw direct analogies to selection-on-observables designs. First, Assumption \ref{assumption_unconfoundedness} holds trivially since conditioning on $X_i$ fully determines $D_i$. Second, Assumption \ref{assumption_common_support} is necessarily violated, as $\propensityScore{x}$ is either zero or one for all $x$ \parencite[see, e.g.,][]{imbens2008regression}.

A standard approach in RD settings is to rely on the following continuity assumption for identification of the Average Treatment Effect at the Cutoff (ATC) $\tau_C := \EX [ \potentialOutcome{1} - \potentialOutcome{0} | X_i = c ]$. 
\renewcommand{\theassumption}{7\alph{assumption}}
\setcounter{assumption}{0}
\begin{assumption}\label{assumption_continuity} 
    (Continuity of mean outcomes): $\EX[ \potentialOutcome{d} | X_i = x ]$ is continuous in $x$ for $d = 0, 1$. 
\end{assumption} 
However, the qualitative nature of $\outcome$ renders $\tau_C$ ill-defined. We thus shift our focus toward the identification of the Probability of Shift at the Cutoff (PSC):
\begin{equation} 
    \probabilityShiftCutoff{m} := \PX ( \potentialOutcome{1} = m | X_i = c) - \PX( \potentialOutcome{0} = m | X_i = c).
\end{equation} 

For identification of PSC, we introduce an alternative continuity assumption.
\begin{assumption}\label{assumption_continuity_probabilities} 
    (Continuity of probability mass functions): $\PX( \potentialOutcome{d} = m | X_i = x )$ is continuous in $x$ for $d = 0, 1$. 
\end{assumption} 
\renewcommand{\theassumption}{\arabic{assumption}} % Restore default numbering
% 
%\noindent Assumption \ref{assumption_continuity_probabilities} requires that the conditional probability mass functions of potential outcomes evolve smoothly with $x$. %This ensures that class probabilities -- rather than the expected outcomes -- remain comparable in a neighborhood of $c$, enabling the identification of $\probabilityShiftCutoff{m}$ through a fair comparison of class probabilities on either side of the threshold.

\noindent Under Assumptions \ref{assumption_unconfoundedness} and \ref{assumption_continuity_probabilities},
\begin{equation}
    \probabilityShiftCutoff{m} = \lim_{x \downarrow c} \PX ( \outcome = m | D_i = 1, X_i = x) - \lim_{x \uparrow c} \PX ( \outcome = m | D_i = 0, X_i = x).
\end{equation}
This establishes that a slight modification of the standard RD assumptions is sufficient to identify the PSC. 

%%% DIFFERENCE-IN-DIFFERENCES.
\subsection{Difference-in-differences} 
\label{sec_difference_in_differences}
 %Difference-in-Differences (DiD) designs are employed to identify causal effects when units are observed over time and treatment is introduced only from a certain point onward for some units. The key assumption is that, in the absence of treatment, the change in outcomes for treated units would have mirrored the change observed in the control group \parencite[see, e.g.,][]{roth2023s, deChaisemartin2023}. 
\noindent We focus on the standard two-period Difference-in-Differences (DiD) framework for ease of illustration. We observe units in two time periods $s \in \{ t, t - 1 \}$ and use $\potentialOutcomeTime{d}{s}$ to index potential outcomes by both treatment status $d$ and time $s$. No units receive treatment in the pre-treatment period $t - 1$, while in the post-treatment period $t$ a subset of units is treated ($D_i = 1$). In this context, the PST is defined as
\begin{equation}
    \probabilityShiftTreated{m} := \PX ( \potentialOutcomeTime{1}{t} = m | D_i = 1 ) - \PX ( \potentialOutcomeTime{0}{t} = m | D_i = 1 ). 
    \label{equation_probability_shift_treated_diff_in_diff}
\end{equation}

A standard approach in DiD settings is to impose the following parallel trends assumption for identification of the ATT, now defined as $\tau_T := \EX [ \potentialOutcomeTime{1}{t} - \potentialOutcomeTime{0}{t} | D_i = 1 ]$.
\begin{altAssumptionEnv} \label{assumption_parallel_trends}
     (Parallel trends on mean outcomes): \\ $\EX [ \potentialOutcomeTime{0}{t} - \potentialOutcomeTime{0}{t-1} | D_i = 1 ] = \EX [ \potentialOutcomeTime{0}{t} - \potentialOutcomeTime{0}{t-1} | D_i = 0 ]$.
\end{altAssumptionEnv}
%
%\noindent Assumption \ref{assumption_parallel_trends} states that, in the absence of treatment, the change in outcomes for the treated group would have mirrored that of the control group, effectively ensuring that any systematic differences between the groups remain constant over time. This allows for a fair comparison of how outcomes evolve across groups, thereby enabling the identification of $\tau_T$.

However, the qualitative nature of $\outcome$ renders $\tau_T$ meaningless, and Assumption \ref{assumption_parallel_trends} does not enable the identification of the PST. We thus introduce an alternative parallel trends assumption.
\begin{altAssumptionEnv}\label{assumption_parallel_trends_probabilities}
    (Parallel trends on probability mass functions): \\ $\PX ( \potentialOutcomeTime{0}{t} = m | D_i = 1) - \PX ( \potentialOutcomeTime{0}{t-1} = m | D_i = 1 ) = \PX ( \potentialOutcomeTime{0}{t} = m | D_i = 0) - \PX ( \potentialOutcomeTime{0}{t-1} = m | D_i = 0 )$.
\end{altAssumptionEnv}
%
%\noindent Assumption \ref{assumption_parallel_trends_probabilities} requires that the probability time shift of $\potentialOutcomeTime{0}{s}$ for class $m$ follows a similar evolution over time in both the treated and control groups. %Essentially, this imposes parallel trends on the probability mass functions of $\potentialOutcomeTime{0}{s}$, rather than on its mean, ensuring that categorical shifts -- rather than differences -- remain comparable across groups.

\noindent Under Assumption \ref{assumption_parallel_trends_probabilities}, we obtain
%     %
%     \begin{equation}
%         \begin{split}
%             \PX ( \potentialOutcomeTime{0}{t} = m | D_i = 1 ) & = \PX ( \potentialOutcomeTime{0}{t-1} = m | D_i = 1 ) + \\
%             %
%             & \PX ( \potentialOutcomeTime{0}{t} = m | D_i = 0) - \PX (\potentialOutcomeTime{0}{t-1} = m | D_i = 0 ).
%         \end{split}
%     \end{equation}
% %
% Replacing this in (\ref{equation_probability_shift_treated_diff_in_diff}) and using (\ref{equation_observational_rule_time}) to replace potential outcomes with observable outcomes, we obtain
%
\begin{equation} 
    \probabilityShiftTreated{m} = \PX ( \outcomeTime{t} = m | D_i = 1 ) - \PX ( \outcomeTime{t-1} = m | D_i = 1 ) - [ \PX ( \outcomeTime{t} = m | D_i = 0) - \PX (\outcomeTime{t-1} = m | D_i = 0 ) ]
    \label{equation_identification_did} 
\end{equation}
This establishes that a slight modification of the standard DiD assumption is sufficient to identify the PST. 

%%% ESTIMATION.
\subsection{Estimation}
\noindent For estimation, researchers can rely on conventional methods while explicitly modeling category probabilities rather than the outcome itself. For instance, note that
$$
    \probabilityShift{m} = \EX[ \EX [ \indicator{\outcome = m} | D_i = 1, \vector X_i ] - \EX [ \indicator{\outcome = m} | D_i = 0, \vector X_i ]].
$$
This suggests that we can construct an estimate $\estimatedProbabilityShift{m}$ by applying standard estimation strategies to the binary variable $\indicator{\outcome = m}$ and estimating the ATE on that variable.\footnote{\ Applying such estimation strategies to the subsample of treated units would yield an estimate $\estimatedProbabilityShiftTreated{m}$ of $\probabilityShiftTreated{m}$.} Similar reasoning applies to other parameters. This approach allows applied researchers to utilize familiar econometric tools while targeting causal parameters that are both meaningful and interpretable.

%%%%%%%%%%%%%%%%%%%%%%%%%%%%%%%%%%%%%%%%
%              CONCLUSION.             %
%%%%%%%%%%%%%%%%%%%%%%%%%%%%%%%%%%%%%%%%

\section{Conclusion} 
\label{sec_conclusion}
\noindent This paper highlights the fundamental challenges that arise when conventional causal inference methods are applied to qualitative outcomes and advocates for a fundamental shift toward alternative estimands that are meaningful, interpretable, and identifiable within standard research designs. Our approach ensures validity and interpretability without departing from widely used research designs.

% While we establish identification and estimation within standard cases for ease of illustration, our framework naturally extends to more complex settings, including instrumental variables with multiple instruments \parencite[see, e.g.,][]{HLM2025,mogstad2018using,mogstad2018identification,goff2024vector}, difference-in-differences with heterogeneous effects and staggered treatment adoption \parencite[see, e.g.,][]{callaway2021difference,SunAbraham2021,Borusyak2024, goodman2021difference}, and fuzzy regression discontinuity designs  \parencite[see, e.g.,][]{cattaneo2022regression}. We plan to continuously update our R package, \texttt{causalQual}, to incorporate these extensions as they develop.\footnote{\ Users interested in specific extensions are encouraged to contribute by submitting feature requests or discussions at \href{https://github.com/riccardo-df/causalQual/issues}{https://github.com/riccardo-df/causalQual/issues}.}

% %%%%%%%%%%%%%%%%%%%%%%%%%%%%%%%%%%%%%%%%
% %       DECLARATION OF INTEREST.       %
% %%%%%%%%%%%%%%%%%%%%%%%%%%%%%%%%%%%%%%%%

% \section*{Declaration of Interest Statement}
% \noindent The authors report there are no competing interests to declare.

% %%%%%%%%%%%%%%%%%%%%%%%%%%%%%%%%%%%%%%%%
% %          DATA AVAILABILITY.          %
% %%%%%%%%%%%%%%%%%%%%%%%%%%%%%%%%%%%%%%%%

% \section*{Data availability statement}
% \noindent Data sharing is not applicable to this article as no new data were created or analyzed in this study.

%%%%%%%%%%%%%%%%%%%%%%%%%%%%%%%%%%%%%%%%
%           GENERATIVE AI.             %
%%%%%%%%%%%%%%%%%%%%%%%%%%%%%%%%%%%%%%%%

\section*{Declaration of generative AI}
\noindent During the preparation of this work the authors used ChatGPT only to improve language and readability. After using this tool, the authors reviewed and edited the content as needed and take full responsibility for the content of the publication.

%%%%%%%%%%%%%%%%%%%%%%%%%%%%%%%%%%%%%%%%
%           ACKNOWELDGMENTS.           %
%%%%%%%%%%%%%%%%%%%%%%%%%%%%%%%%%%%%%%%%

\section*{Acknowledgements}
\noindent We are grateful to Jana Mareckova, Michael Lechner, and Franco Peracchi for useful comments and discussions. The R package for implementing the methodologies developed in this paper is available on CRAN at \href{https://cran.r-project.org/web/packages/causalQual/index.html}{https://cran.r-project.org/web/packages/causalQual/index.html}. The associated website is at \href{https://riccardo-df.github.io/causalQual/}{https://riccardo-df.github.io/causalQual/}.

%%%%%%%%%%%%%%%%%%%%%%%%%%%%%%%%%%%%%%%%
%             REFERENCES.              %
%%%%%%%%%%%%%%%%%%%%%%%%%%%%%%%%%%%%%%%%

%\singlespacing
\doublespacing
\newrefcontext[sorting = nty]
\newpage
\printbibliography

% %%%%%%%%%%%%%%%%%%%%%%%%%%%%%%%%%%%%%%%%
% %              APPENDIX.               %
% %%%%%%%%%%%%%%%%%%%%%%%%%%%%%%%%%%%%%%%%

\newpage
\appendix
\doublespacing

\setcounter{section}{0}
\renewcommand{\thesection}{\Alph{section}}

\setcounter{equation}{0}
\renewcommand{\theequation}{\thesection.\arabic{equation}}

\setcounter{assumption}{0}
\renewcommand{\theassumption}{\thesection.\arabic{assumption}}

\setcounter{note}{0}
\renewcommand{\thenote}{\thesection.\arabic{note}}

\setcounter{remark}{0}
\renewcommand{\theremark}{\thesection.\arabic{remark}}

\setcounter{table}{0}
\renewcommand{\thetable}{\thesection.\arabic{table}}

\setcounter{figure}{0}
\renewcommand{\thefigure}{\thesection.\arabic{figure}}

%%% PROOFS.
\section{Proofs}
\label{app_proofs}

\begin{proof}[Identification of PS, PST, and CPS under selection-on-observables]
    First, we show that Assumptions \ref{assumption_unconfoundedness}--\ref{assumption_common_support} suffice for identifying the CPS at $x$:
    \begin{equation*}
        \begin{split}
            \conditionalClassProbability{m}{1}{\vector x} & - \conditionalClassProbability{m}{0}{\vector x} = \\
            & = \PX ( \potentialOutcome{1} = m | D_i = 1, \vector X_i = \vector x ) - \PX ( \potentialOutcome{0} = m | D_i = 0, \vector X_i = \vector x ) \,\,\,\,\,\,\, (\textit{by definition of $\outcome$}) \\
            & = \PX ( \potentialOutcome{1} = m | \vector X_i = \vector x ) - \PX ( \potentialOutcome{0} = m | \vector X_i = \vector x )\,\,\,\,\,\,\,\,\,\,\,\,\,\,\,\,\,\,\,\,\,\,\,\,\,\,\,\,\,\,\,\,\,\,\,\,\,\,\,\,\,\,\,\,\, (\textit{by Assumption \ref{assumption_unconfoundedness}}) \\
            & = \conditionalProbabilityShift{m}{\vector x},
        \end{split}
    \end{equation*}
    where Assumption \ref{assumption_common_support} ensures that all conditional probabilities are well-defined at $\vector x$. Since $\probabilityShift{m} = \EX [ \conditionalProbabilityShift{m}{\vector X_i} ]$ and $\probabilityShiftTreated{m} = \EX [ \conditionalProbabilityShift{m}{\vector X_i} | D_i = 1 ]$, the PS and PST are also identified.  
\end{proof}

\begin{proof}[Identification of LPS under IV]
    By the Law of Total Probability,
    \begin{eqnarray*}
        \PX (\outcome = m | Z_i = z )
        & = & \PX(\outcome = m | D_i=1, Z_i = z) \PX(D_i=1 | Z_i =z)
        \notag\\
        && \quad + \PX(\outcome = m | D_i=0, Z_i=z) \PX(D_i=0 | Z_i=z).
        \label{z1}
    \end{eqnarray*}
    Therefore,
    \begin{equation}
        \begin{split}
        \PX (\outcome = m & | Z_i = 1 ) - \PX (\outcome = m | Z_i = 0 ) = \\
        & \PX(\outcome = m | D_i=1, Z_i = 1) \PX(D_i=1 | Z_i=1) + \PX(\outcome = m | D_i=0, Z_i=1) \PX(D_i=0 | Z_i=1) \\
        - & \PX(\outcome = m | D_i=1, Z_i = 0) \PX(D_i=1 | Z_i=0) - \PX(\outcome = m | D_i=0, Z_i=0) \PX(D_i=0 | Z_i=0).
        \end{split}
        \label{equation_this}
    \end{equation}
    By Assumption \ref{assumption_monotonicity}, $\PX (T_i = de) = 0$, thus
    \begin{equation}
        \PX(D_i=1 | Z_i=0) = \PX(T_i=at), \quad \PX(D_i=0 | Z_i=1) = \PX(T_i=nt),
        \label{shares}
    \end{equation}
    where Assumption \ref{assumption_exogeneity_iv} ensures that $\PX ( T_i = t | Z_i ) = \PX ( T_i = t )$. Therefore, (\ref{equation_this}) becomes
    \begin{equation}
        \begin{split}
        \PX (\outcome = m & | Z_i = 1 ) - \PX (\outcome = m | Z_i = 0 ) = \\
        & \PX(\outcome = m | D_i=1, Z_i = 1) \PX(D_i=1 | Z_i=1) + \PX(\outcome = m | D_i=0, Z_i=1) \PX(T_i=nt) \\
        - & \PX(\outcome = m | D_i=1, Z_i = 0) \PX(T_i=at) - \PX(\outcome = m | D_i=0, Z_i=0) \PX(D_i=0 | Z_i=0).
        \end{split}
        \label{equation_this_other}
    \end{equation}
    
    By Assumption \ref{assumption_exogeneity_iv}, and using the observational rule and Assumption \ref{assumption_exclusion} to replace observable outcomes with potential outcomes, we obtain
    \begin{equation*}
        \PX (\outcome = m | D_i=1, Z_i=1) = \frac{\PX(T_i=co)\PX(\potentialOutcome{1} = m | T_i=co) + \PX(T_i=at) \PX(\potentialOutcome{1} = m | T_i=at)}{\PX(D_i=1 | Z_i=1)},
    \end{equation*}
    and
    \begin{equation*}
        \PX(\outcome = m | D_i=0, Z_i=0) =\frac{\PX(T_i=co) \PX(\potentialOutcome{0} = m | T_i=co)+\PX(T_i=nt) \PX(\potentialOutcome{0} = m | T_i=nt)}{\PX(D_i=0 | Z_i=0)},
    \end{equation*}
    Moreover, by Assumption \ref{assumption_monotonicity}, we have
    \begin{equation*}
        \PX(\outcome = m | D_i=1, Z_i=0) = \PX(\potentialOutcome{1} = m | T_i=at)
    \end{equation*}
    and
    \begin{equation*}
        \PX(\outcome = m | D_i=0, Z_i=1) = \PX(\potentialOutcome{0} = m | T_i=nt).
    \end{equation*}
    Therefore, (\ref{equation_this_other}) simplifies to
    \begin{equation*}
        \PX (\outcome = m | Z_i = 1 ) - \PX (\outcome = m | Z_i = 0 ) = \PX(T_i=co) \LocalProbabilityShift{m}.
    \end{equation*}
    
    Finally, using previous results and Assumption \ref{assumption_relevance},
    \begin{equation*}
        \PX(T_i=co) = \PX(D_i=1 | Z_i=1) - \PX(D_i=1 | Z_i=0) > 0.
    \end{equation*}
    which implies
    \begin{equation*}
        \frac{\PX (\outcome = m | Z_i = 1 ) - \PX (\outcome = m | Z_i = 0 )}{\PX(D_i=1 | Z_i=1) - \PX(D_i=1 | Z_i=0)} = \LocalProbabilityShift{m}.
    \end{equation*}
\end{proof}

\begin{proof}[Identification of PSC under RD]
    Under Assumptions \ref{assumption_unconfoundedness} and \ref{assumption_continuity_probabilities}, we obtain
    \begin{equation*}
        \begin{split}
            \lim_{x \downarrow c} & \PX ( \outcome = m | D_i = 1, X_i = x) =  \\
            & = \lim_{x \downarrow c} \PX ( \potentialOutcome{1} = m | D_i = 1, X_i = x) \,\,\,\,\,\,\, (\textit{by defintion of $\outcome$}) \\
            & = \lim_{x \downarrow c} \PX ( \potentialOutcome{1} = m | X_i = x)\,\,\,\,\,\,\,\,\,\,\,\,\,\,\,\,\,\,\,\,\,\,\,\,\,\, (\textit{by Assumption \ref{assumption_unconfoundedness}}) \\
            & = \PX ( \potentialOutcome{1} = m | X_i = c), \,\,\,\,\,\,\,\,\,\,\,\,\,\,\,\,\,\,\,\,\,\,\,\,\,\,\,\,\,\,\,\,\, (\textit{by Assumption \ref{assumption_continuity_probabilities}})
        \end{split}
        \label{equation_identification_rd}
    \end{equation*}
    and similarly
    \begin{equation*}
        \lim_{x \uparrow c} \PX ( \outcome = m | D_i = 0, X_i = x) = \PX ( \potentialOutcome{0} = m | X_i = c).
    \end{equation*}
    Thus,
    \begin{equation}
        \lim_{x \downarrow c} \PX ( \outcome = m | D_i = 1, X_i = x) - \lim_{x \uparrow c} \PX ( \outcome = m | D_i = 0, X_i = x) = \probabilityShiftCutoff{m}.
    \end{equation}
\end{proof}

\begin{proof}[Identification of PST under DiD]
    Under Assumption \ref{assumption_parallel_trends_probabilities}, we obtain
        \begin{equation}
            \begin{split}
                \PX ( \potentialOutcomeTime{0}{t} = m | D_i = 1 ) & = \PX ( \potentialOutcomeTime{0}{t-1} = m | D_i = 1 ) + \\
                & \PX ( \potentialOutcomeTime{0}{t} = m | D_i = 0) - \PX (\potentialOutcomeTime{0}{t-1} = m | D_i = 0 ).
            \end{split}
        \end{equation}
    Replacing this in (\ref{equation_probability_shift_treated_diff_in_diff}) and using the standard observational rule
    $$
       \outcomeTime{t-1} = \potentialOutcomeTime{0}{t-1}, \quad \outcomeTime{t} = D_i \potentialOutcomeTime{1}{t} + (1 - D_i) \potentialOutcomeTime{0}{t}
    $$
    to replace potential outcomes with observable outcomes, we obtain
    \begin{equation} 
        \begin{split}
            \probabilityShiftTreated{m} = \PX ( \outcomeTime{t} = m | D_i = 1 ) & - \PX ( \outcomeTime{t-1} = m | D_i = 1 ) - \\ 
            [ \PX ( \outcomeTime{t} = m | D_i = 0) & - \PX (\outcomeTime{t-1} = m | D_i = 0 ) ].
        \end{split}
    \end{equation}
\end{proof}

%%%%%%%%%%%%%%%%%%%%%%%%%%%%%%%%%%%%%%%%
%             SIMULATION.              %
%%%%%%%%%%%%%%%%%%%%%%%%%%%%%%%%%%%%%%%%

\section{Simulations} 
\label{app_simulation}
%%% SELECTION-ON-OBSERVABLES.
\subsection{Selection-on-observables}
\label{subsec_simulation_selection_on_observables}
\noindent Potential outcomes are modeled using two distinct approaches, depending on whether the outcome is multinomial or ordered. 

For multinomial outcomes, we construct conditional probability mass functions by first computing linear predictors based on the covariates:
$$
    \eta_{mi} (d) = \beta_{m1}^d X_{i1} + \beta_{m2}^d X_{i2} + \beta_{m3}^d X_{i3}, \quad d= 0, 1,
$$
and then transforming these predictors into valid probability distributions using the logit transformation:
$$
    \PX ( \potentialOutcome{d} = m | \vector X_i ) = \frac{\exp(\eta_{mi} (d))}{\sum_{m'} \exp(\eta_{m'i}(d))}, \quad d = 0, 1.
$$
The coefficients used in the linear predictor models are specified as:
\[
\beta_1 =
\begin{bmatrix}
  0.5  &  0.3  & -0.2  \\ 
 -0.2  &  0.4  &  0.1  \\ 
  0.1  & -0.3  &  0.5  
\end{bmatrix},
\quad
\beta_0 =
\begin{bmatrix}
  0.7  &  0.7  & -0.2  \\ 
 -0.2  &  0.4  &  0.1  \\ 
 -0.2  & -0.5  &  0.1  
\end{bmatrix}.
\]

For ordered outcomes, we first generate underlying latent potential outcomes
$$
   Y_i^* (d) = \tau d + \sum_j X_{ij} + \mathcal{N} (0, 1), \quad d = 0, 1,
$$
and then obtain the ordered potential outcomes by discretizing $Y_i^* (d)$ using fixed thresholds:
$$
    \potentialOutcome{d} = \sum_{m = 1}^M m \indicator{\zeta_{m - 1} < Y_i^* (d) \leq \zeta_m}, \quad d = 0, 1,
$$
with $\zeta_1 = 2$, $\zeta_2 = 3$, and $\tau = 2$.

% For multinomial outcomes, we construct the probability mass functions by applying a logit transformation to linear predictors. Specifically, for each class $m$, we first compute linear predictors based on the covariates and then transform these predictors into valid probability distributions using the logit function. The potential outcomes are then generated by sampling from the discrete set $\{ 1, 2, 3 \}$ according to these mass functions. For ordered outcomes, we first generate underlying latent potential outcomes $Y_i^* (d)$ and then obtain the observable potential outcomes $\potentialOutcome{d}$ by discretizing $Y_i^* (d)$ using fixed thresholds.

We generate three covariates, each independently drawn from a uniform distribution over $[0, 1]$. Treatment is then assigned via a Bernoulli trial $D_i \sim \textit{Bernoulli} ( \propensityScore{\vector X_i} )$. We consider two assignment mechanisms: a randomized design, with $\propensityScore{\vector X_i} = 0.5$, and an observational design, with $\propensityScore{\vector X_i} = (\vector X_{i1} + \vector X_{i3} ) / 2$.

Estimation is performed by applying the double machine learning procedures of \textcite{chernozhukov2018double} to the binary variable $\indicator{\outcome = m}$. Specifically, for each class $m$, we define the doubly robust scores as
\begin{equation}
    \drScores{m} := \conditionalClassProbability{m}{1}{\vector X_i} - \conditionalClassProbability{m}{0}{\vector X_i} + D_i \frac{\indicator{\outcome = m} - \conditionalClassProbability{m}{1}{\vector X_i}}{\propensityScore{\vector X_i}} - (1 - D_i) \frac{\indicator{\outcome = m} - \conditionalClassProbability{m}{0}{\vector X_i}}{1 - \propensityScore{\vector X_i}}.
    \label{equation_dr_scores}
\end{equation}
We then construct plug-in estimates $\estimatedDrScores{m}$ of $\drScores{m}$ by replacing the unknown $\conditionalClassProbability{m}{1}{\cdot}$, $\conditionalClassProbability{m}{0}{\cdot}$, and $\propensityScore{\cdot}$ with consistent estimates $\estimatedConditionalClassProbability{m}{1}{\cdot}$, $\estimatedConditionalClassProbability{m}{0}{\cdot}$, and  $\estimatedPropensityScore{\cdot}$ obtained via $K$-fold cross fitting.\footnote{\ A potential issue in estimation arises if, after splitting the sample into folds and each fold into treated and control groups, at least one class of $\outcome$ is unobserved within a given fold for a specific treatment group. To mitigate this issue, our R package \texttt{causalQual} repeatedly splits the data until all outcome categories are present in each fold and treatment group.} The estimator for PS is then
\begin{equation} 
    \estimatedProbabilityShift{m} = \frac{1}{n} \sum_{i=1}^{n} \estimatedDrScores{m}, 
    \label{equation_estimator_soo}
\end{equation}
and its variance is estimated as
\begin{equation} 
    \widehat{\text{Var}} ( \estimatedProbabilityShift{m} ) = \frac{1}{n} \sum_{i=1}^{n} ( \estimatedDrScores{m} - \estimatedProbabilityShift{m} )^2. 
    \label{equation_estimator_variance_soo}
\end{equation}
We use these estimates to construct confidence intervals using conventional normal approximations.\footnote{\ While the true parameter lies within the interval $[-1,1]$, we do not impose this restriction in constructing our confidence intervals. In principle, leveraging this known bound could sharpen the intervals.}

To estimate the conditional class probabilities $\conditionalClassProbability{m}{d}{\cdot}, \, d = 0, 1$, we adopt multinomial machine learning strategies when the outcome is multinomial and the honest ordered correlation forest estimator when the outcome is ordered \parencite{di2025ordered}. We train separate models for treated and control units. The propensity score $\propensityScore{\cdot}$ is estimated via an honest regression forest \parencite{athey2019generalized}. All these functions are estimated via $5$-fold cross fitting.

% \footnote{\ The choice of estimator for $\estimatedConditionalClassProbability{m}{1}{\cdot}$ and $\estimatedConditionalClassProbability{m}{0}{\cdot}$ should vary depending on whether $\outcome$ is multinomial or ordered, as different methods are better suited to each case \parencite[see, e.g.,][]{di2025ordered}.}
Table \ref{table_simulation_results_selection_on_observables} reports results obtained with $2000$ replications and fixing the sample size at $2000$.
\begin{table}[htbp]
    \centering
    \caption{Simulation results: selection-on-observables.}
    \label{table_simulation_results_selection_on_observables}
    \begin{threeparttable}
        \begin{tabular}{l cccc cccc}
            \toprule
            & \multicolumn{4}{c}{\textit{Multinomial}} & \multicolumn{4}{c}{\textit{Ordered}} \\
            \cmidrule(lr){2-5} \cmidrule(lr){6-9}
            & {$\probabilityShift{m}$} & {$|Bias|$} & {$SD$} & {$95\%$}
            & {$\probabilityShift{m}$} & {$|Bias|$} & {$SD$} & {$95\%$} \\
            \midrule
            \multicolumn{9}{l}{\textbf{Panel A: Randomized design}} \\
            Class 1 & 0.00 & 0.00 & 0.02 & 0.95 & -0.58 & 0.00 & 0.02 & 0.95 \\
            Class 2 & -0.05 & 0.00 & 0.02 & 0.95 & 0.00 & 0.00 & 0.02 & 0.95 \\
            Class 3 & 0.05 & 0.00 & 0.02 & 0.95 & 0.58 & 0.00 & 0.02 & 0.93 \\
            \midrule
            \multicolumn{9}{l}{\textbf{Panel B: Observational design}} \\
            Class 1 & 0.00 & 0.00 & 0.02 & 0.94 & -0.58 & 0.01 & 0.02 & 0.93 \\
            Class 2 & -0.05 & 0.00 & 0.03 & 0.93 & 0.00 & 0.00 & 0.02 & 0.95 \\
            Class 3 & 0.05 & 0.00 & 0.02 & 0.94 & 0.58 & 0.01 & 0.02 & 0.93 \\
            \bottomrule
        \end{tabular}
    \end{threeparttable}
\end{table}

\noindent All estimates are approximately unbiased and coverage rates are close to the nominal level.

%%% INSTRUMENTAL VARIABLES.
\subsection{Instrumental variables}
\label{subsec_simulation_iv}
\noindent Potential outcomes and covariates are generated following the procedure of the previous subsection. We further generate a binary instrument $Z_i \sim \textit{Bernoulli}(0.5)$. Treatment is assigned as a Bernoulli trial $D_i \sim \textit{Bernoulli} ( \propensityScore{\vector X_i, Z_i} )$, where we allow the instrument to influence treatment probability. We only consider an observational setting, with $\propensityScore{\vector X_i, Z_i} = (\vector X_{i1} + \vector X_{i2} + Z_i ) / 3$. However, we assume that the covariates $\vector X_i$ are unobserved, preventing their use in adjusting for selection into treatment.

Estimation is carried out by applying the standard two-stage least squares method to the binary variable $\indicator{\outcome = m}$. Specifically, we first estimate the following first-stage regression model via OLS:
\begin{equation}
    D_i = \gamma_0 + \gamma_1 Z_i + \nu_i,
\end{equation}
and construct the predicted values $\widehat{D}_i$. We then use these predicted values in the second-stage regressions:
\begin{equation}
    \indicator{\outcome = m} = \alpha_{m0} + \alpha_{m1} \widehat{D}_i + \epsilon_{mi}, \quad m = 1, \dots M.
    \label{equation_second_stage_iv}
\end{equation}
A well-established result in the IV literature is that, under Assumptions \ref{assumption_exogeneity_iv}--\ref{assumption_relevance}, $\alpha_{m1} = \LocalProbabilityShift{m}$ 
\parencite{imbens1994identification}. Therefore, we estimate \eqref{equation_second_stage_iv} via OLS and use the resulting estimate $\hat{\alpha}_{m1}$ as an estimate of $\LocalProbabilityShift{m}$. \begin{table}[b]
    \centering
    \caption{Simulation results: instrumental variables.}
    \label{table_simulation_results_iv}
    \begin{threeparttable}
        \begin{tabular}{l cccc cccc}
            \toprule
            & \multicolumn{4}{c}{\textit{Multinomial}} & \multicolumn{4}{c}{\textit{Ordered}} \\
            \cmidrule(lr){2-5} \cmidrule(lr){6-9}
            & {$\LocalProbabilityShift{m}$} & {$|Bias|$} & {$SD$} & {$95\%$}
            & {$\LocalProbabilityShift{m}$} & {$|Bias|$} & {$SD$} & {$95\%$} \\
            \midrule
            Class 1 & 0.00 & 0.00 & 0.06 & 0.95 & -0.58 & 0.00 & 0.05 & 0.95 \\
            Class 2 & -0.05 & 0.00 & 0.06 & 0.95 & 0.00 & 0.00 & 0.06 & 0.96 \\
            Class 3 & 0.05 & 0.00 & 0.06 & 0.95 & 0.58 & 0.00 & 0.05 & 0.95 \\
            \bottomrule
        \end{tabular}
    \end{threeparttable}
\end{table}
 
Standard errors are computed using standard procedures and used to construct conventional confidence intervals.\footnote{\ Our R package \texttt{causalQual} implements this two-stage least squares procedure by calling the \texttt{ivreg} function from the R package \texttt{AER}.}

Table \ref{table_simulation_results_iv} reports results obtained with $10000$ replications and fixing the sample size at $2000$. All estimates are approximately unbiased and coverage rates are close to the nominal level.

%%% REGRESSION DISCONTINUITY.
\subsection{Regression discontinuity}
\label{subsec_simulation_rd}
\noindent Potential outcomes and covariates are generated following the procedure of the previous sections. We designate $X_{i1}$ as the running variable, with treatment assigned to units such as $X_{i1} \geq 0.5$. Thus, treatment status is determined as $D_i = \indicator{X_{i1} \geq 0.5}$.

Estimation is performed using standard local polynomial estimators applied to the binary variable $\indicator{\outcome = m}$. Specifically, we implement the robust bias-corrected inference procedure of \textcite{calonico2014robust}.\footnote{\ Our R package \texttt{causalQual} implements these estimation strategies by calling the \texttt{rdrobust} function from the R package \texttt{rdrobust} \parencite{calonico2015rdrobust}.}

Table \ref{table_simulation_results_rd} reports results obtained with $10000$ replications and fixing the sample size at $2000$. All estimates are approximately unbiased and coverage rates are close to the nominal level.

\begin{table}[htbp]
    \centering
    \caption{Simulation results: regression discontinuity.}
    \label{table_simulation_results_rd}
    \begin{threeparttable}
        \begin{tabular}{l cccc cccc}
            \toprule
            & \multicolumn{4}{c}{\textit{Multinomial}} & \multicolumn{4}{c}{\textit{Ordered}} \\
            \cmidrule(lr){2-5} \cmidrule(lr){6-9}
            & {$\probabilityShiftCutoff{m}$} & {$|Bias|$} & {$SD$} & {$95\%$}
            & {$\probabilityShiftCutoff{m}$} & {$|Bias|$} & {$SD$} & {$95\%$} \\
            \midrule
            Class 1 & 0.00 & 0.00 & 0.10 & 0.94 & -0.60 & 0.00 & 0.08 & 0.94 \\
            Class 2 & -0.05 & 0.00 & 0.10 & 0.95 & 0.00 & 0.00 & 0.09 & 0.94 \\
            Class 3 & 0.05 & 0.00 & 0.10 & 0.94 & 0.60 & 0.00 & 0.09 & 0.93 \\
            \bottomrule
        \end{tabular}
    \end{threeparttable}
\end{table}

%%% DIFFERENCE-IN-DIFFERENCES.
\subsection{Difference-in-differences}
\label{subsec_simulation_diff_in_diff}
\noindent Potential outcomes are generated following the procedure of the previous sections, applying the same models to both time periods---essentially treating time as having no direct effect on potential outcomes. Estimation is performed by applying the canonical two-group/two-period DiD method to the binary variable $\indicator{\outcome = m}$. Specifically, consider the following linear regression model:
\begin{equation}
    \indicator{\outcomeTime{s} = m} = D_i \beta_{m1} + \indicator{s = t} \beta_{m2} + D_i \indicator{s = t} \beta_{m3} + \epsilon_{mis}.
    \label{equation_diff_in_diff_linear_model}
\end{equation}
A well-established result in the DiD literature is that, under Assumption \ref{assumption_parallel_trends_probabilities}, $\beta_{m3} = \probabilityShiftTreated{m}$ \parencite[see, e.g.,][]{goodman2021difference}. Therefore, we estimate model \eqref{equation_diff_in_diff_linear_model} via OLS and use the resulting estimate $\hat{\beta}_{m3}$ as an estimate of $\probabilityShiftTreated{m}$. Standard errors are clustered at the unit level and used to construct conventional confidence intervals.

Table \ref{table_simulation_results_did} reports results obtained with $10000$ replications and fixing the sample size at $2000$. All estimates are approximately unbiased and coverage rates are close to the nominal level.

\begin{table}[htbp]
    \centering
    \caption{Simulation results: difference-in-differences.}
    \label{table_simulation_results_did}
    \begin{threeparttable}
        \begin{tabular}{l cccc cccc}
            \toprule
            & \multicolumn{4}{c}{\textit{Multinomial}} & \multicolumn{4}{c}{\textit{Ordered}} \\ \cmidrule(lr){2-5} \cmidrule(lr){6-9}
            & {$\probabilityShiftTreated{m}$} & {$|Bias|$} & {$SD$} & {$95\%$} & {$\probabilityShiftTreated{m}$} & {$|Bias|$} & {$SD$} & {$95\%$} \\
            \midrule
            % \multicolumn{9}{l}{\textbf{Panel A: Randomized design}} \\
            % Class 1 & 0.00 & 0.00 & 0.03 & 0.95 & -0.58 & 0.00 & 0.03 & 0.95 \\
            % Class 2 & -0.05 & 0.00 & 0.03 & 0.95 & 0.00 & 0.00 & 0.03 & 0.95 \\
            % Class 3 & 0.05 & 0.00 & 0.03 & 0.95 & 0.58 & 0.00 & 0.02 & 0.95 \\
            % \midrule
            % \multicolumn{9}{l}{\textbf{Panel B: Observational design}} \\
            Class 1 & 0.00 & 0.00 & 0.03 & 0.95 & -0.55 & 0.00 & 0.02 & 0.95 \\
            Class 2 & -0.06 & 0.00 & 0.03 & 0.95 & -0.06 & 0.00 & 0.03 & 0.96 \\
            Class 3 & 0.04 & 0.00 & 0.03 & 0.95 & 0.61 & 0.00 & 0.02 & 0.95 \\
            \bottomrule
        \end{tabular}
    \end{threeparttable}
\end{table}

\end{document}